\documentclass{jkas}

\def\year{2018} 
\def\month{April} 
\def\volume{00} 
\def\issue{0} 
\def\beginpage{1} 
\def\endpage{99} 
\def\received{---} 
\def\accepted{---} 
\date{Received \received; accepted \accepted}

\title{
Two-Component Advective Flows around Neutron Stars
}

\author[1]{Ayan Bhattacharjee}
\author[2]{Sandip K. Chakrabarti}

\affil[1]{S. N. Bose National Centre for Basic Sciences, Salt Lake, Kol: 700106, India; \email{ayan12@bose.res.in}}
\affil[2]{Indian Center for Space Physics, 43 Chalantika, Garia St. Road, Kol: 700084, India; \email{sandipchakrabarti9@gmail.com}}

\def\corrauthor{
	Ayan Bhattacharjee
}

\def\runningauthor{
Bhattacharjee and Chakrabarti
}

\def\runningtitle{
TCAF around Neutron Stars
}

\def\keywords{
radiation: dynamics --- stars: neutron --- accretion, accretion discs --- shock waves
}

\def\abstracttext{
The fundamental difference between accretion around black holes and neutron stars is the inner boundary condition, which affects the behavior of matter very close to the compact objects. This leads to formation of additional shocks and boundary layers for neutron stars. Previous studies on the formation of such boundary layers focused on Keplerian flows that reached the surface of the star, either directly or through the formation of a transition layer. However, behavior of sub-Keplerian matter near the surface of a neutron star has not been studied in detail. Here, we study the effect of viscosity, in presence of cooling, on the sub-Keplerian flows around neutron stars, using Smoothed Particle Hydrodynamics. Our time-dependent study shows that multiple shocks, transition and boundary layers form in such type of accretion, when viscosity is significant, and one or more layers are absent when the viscosity is moderate. These flows are particularly of interest for the wind dominated systems such as Cir X-1. We also report the formation of a generalized flow configuration, Two-Component Advective Flow, for the first time.
}

\begin{document}
\jkashead 


\section{Introduction}
The problem of accretion of rotating matter onto weakly magnetic neutron stars can be divided into three broad approaches. In Titarchuk, Lapidus and Muslimov (1998), the formation of a transition layer around a neutron star was studied in detail. The authors showed that if matter has to reach the star from a Keplerian disk, it has to match the angular velocity of the star at the inner boundary. As the rotation of the star is always sub-Keplerian in nature, it meant that a Keplerian to sub-Keplerian `transition' layer is formed to accommodate that effect, in the presence of viscosity. Study of spreading boundary layers on the surface assumed that the flow remains Keplerian till it reaches the surface and undergoes readjustment on the surface itself through a thin boundary layer (Inogamov and Sunyaev 1999). Both approaches provide valuable insight onto Keplerian to sub-Keplerian angular momentum transitions at or near the surface, but assume that the inflow is Keplerian to begin with, which is still used in recent studies of spreading layers (Abolmasov, N{\"a}ttil{\"a}, \& Poutanen 2020). The third approach allows readjustment of the angular momentum of the flow and the spin of star at the surface of the star in the accretion phase, but also assumes the inflow to be a Keplerian disk (Bhattacharyya and Chakrabarty 2017). In this paper, we look into the common assumption of all three above-mentioned approaches, i.e. a Keplerian inflow, and change it by starting with sub-Keplerian flows. For such flows with small but non-zero angular momenta, in presence of cooling, we study the effect of viscosity in the creation of different types of disks, transition and boundary layers. 
In Bhattacharjee \& Chakrabarti 2019 (hereafter Paper I), the study of accretion flows around weakly magnetic neutron stars is divided into two broad classes: 1. The flow is inviscid, advective and has a low efficiency of radiation. It also comprises of winds from the companion star. 2. The flow has a significant viscosity to redistribute the specific angular momenta. The authors discussed the Class 1 flows in Paper I, i.e., the effects of angular momentum of a sub-Keplerian flow in absence of significant dynamic viscosity. The less-explored domain of sub-Eddington, inviscid accretion around neutron stars (NSs) was studied in detail using hydrodynamic simulations. Paper I also argues why a Two-Component Advective Flow or TCAF (Chakrabarti 1995, 1997; Chakrabarti and Titarchuk 1995) solution is most likely the generalized solution for accreting matter around a weakly magnetic neutron star as well. For a detailed review of the studies, refer to Bhattacharjee 2018, and the references of Paper I. The TCAF solution has been proved to be a stable configuration through numerical simulations (Giri and Chakrabarti 2013) and have been used to explain the spectral and timing properties of stellar mass black holes in a self-consistent manner for multiple sources (Debnath et al. 2014, Bhattacharjee et al. 2017, Banerjee et al. 2019, Banerjee et al. 2020). The paradigm can also be applied to NSs with some modifications (Bhattacharjee \& Chakrabarti 2017, hereafter BC17; Bhattacharjee and Chakrabarti 2020). For inviscid flows with sub-Keplerian angular momentum, two axi-symmetric shocks are present in an accretion flow. The outer shock is the so-called CENtrifugal pressure dominated BOundary Layer or CENBOL (Chakrabarti 1995, 1997) and the inner one is the Normal BOundary Layer (NBOL; BC17, Chakrabarti 2017, Bhattacharjee 2018, Paper I). In the present paper, we focus on the effects of the introduction of viscosity and its gradual enhancement on the flow configuration from a theoretical point-of-view.

\section{Method}
The Smoothed Particle Hydrodynamics (SPH) method was introduced by Monaghan (1992). We use the SPH code used in Paper I (based on the original code of Molteni, Sponholtz and Chakrabarti 1996) to solve for the conservation of mass, momentum and energy. All the equations, notations and boundary conditions (inner and outer) are kept the same. We add the $\alpha$-viscosity prescription for the system following Lanzafame, Molteni and Chakrabarti (1998; hereafter, LMC98), which modifies the following.

1. Conservation of azimuthal component of the momentum:
\begin{equation}
\footnotesize{\Big{(} \frac{D v_{\phi}}{Dt} \Big{)} = - \Big{(} \frac{v_{\phi}v_r}{r} \Big{)} + \frac{1}{\rho}\Big{[} \frac{1}{r^2} \frac{\partial}{\partial r} (r^2 \tau_{r\phi}) \Big{]},}
\end{equation}
where, $\tau_{r\phi}=\mu r \frac{\partial \Omega}{\partial r}$, $\Omega=\frac{v_{\phi}}{r}$. Here, we use the standard Shakura-Sunyaev turbulent viscosity (Shakura \& Sunyaev 1973) $\mu=\alpha \rho a Z_{disk}$, where $Z_{disk}$ is the vertical thickness of the flow as obtained from the vertical equilibrium condition and is given by $Z^2_{disk}=\frac{2}{\gamma} a r (r-1)^2$ (Chakrabarti 1989, 1990; LMC98).

2. Conservation of energy (viscous heating term added)
\begin{equation}
\frac{D}{Dt}\Big{(}e+\frac{1}{2}\vec{v}^2\Big{)}=-\frac{P}{\rho} \vec{\nabla} \cdot \vec{v} + \vec{v} \cdot \Big{(}\frac{D\vec{v}}{Dt}\Big{)} - \zeta_{1/2} \rho e^{\alpha} + \frac{\mu}{\rho}\Big{[}r \frac{\partial \Omega}{\partial r}  \Big{]}^2
\end{equation}
Equations 1 and 2 are identical to the Eq. 15 and 6 of Paper I, apart from the last terms which introduce the viscous effects in the flow (in Paper I, $\mu$ was 0). The flow was injected from $r_{inj}=30r_S$ with $v=0.1211$, $a=0.0590$ and $\lambda_{inj}=1.7$ (similar to case C2 of Paper I).

\section{Results}
\subsection{Formation of Boundary and Transition Layers}
We increased the viscosity parameter from $0.075$ (C1) to $0.15$ (C2) to $0.3$ (C3) and kept injected $\lambda_{inj}=1.7$ (see, Table 1). We also define the RAdiative KEplerian Disk as the equivalent of a standard Keplerian Disk following the prescription of Chakrabarti \& Sahu 1997 (hereafter, CS97), when the effect of the radiative pressure term due to the emission from NBOL is included. If the average repulsive radiative force is $F_{rad}=\frac{<C>}{2(R-1)^2}$, then the effective gravitational force reduces to $F_{g}=\frac{1-<C>}{2(R-1)^2}$ (CS97, Paper I).

\textbf{Case C1:} In presence of lower viscosity ($\alpha=0.075$), the sub-Keplerian flow adjusted its angular momentum very close to $R_{NS}$. The layer where most of this transition took place was the previously identified NBOL. In a very small region ($5-7~r_S$), a RAdiative KEplerian Disk (RAKED) appears and disappears from time to time. Figure 4 shows one such instance when the RAKED is formed. Rest of the outer flow remained sub-Keplerian in nature. This is similar to the cases reported in Paper I. We also see that the turbulent nature of the inner flow, which shows multiple bending instabilities, is very similar to the cases reported in Paper I. Furthermore, matter is ejected from both CENBOL and NBOL in the upper quadrant and only from CENBOL in the lower quadrant. The outflowing matter from NBOL undergoes multiple shock transitions (one before merging with CENBOL-outflow, one after), before becoming transonic near the outer edge. A part of the ejected matter falls back to CENBOL and another part is accreted onto the NS through more radial shocks. This fallback onto the star is primarily achieved due to the lowered angular momentum near the surface.
\begin{figure}[h]
\centering
\includegraphics[height=5.0cm,width=3.0cm]{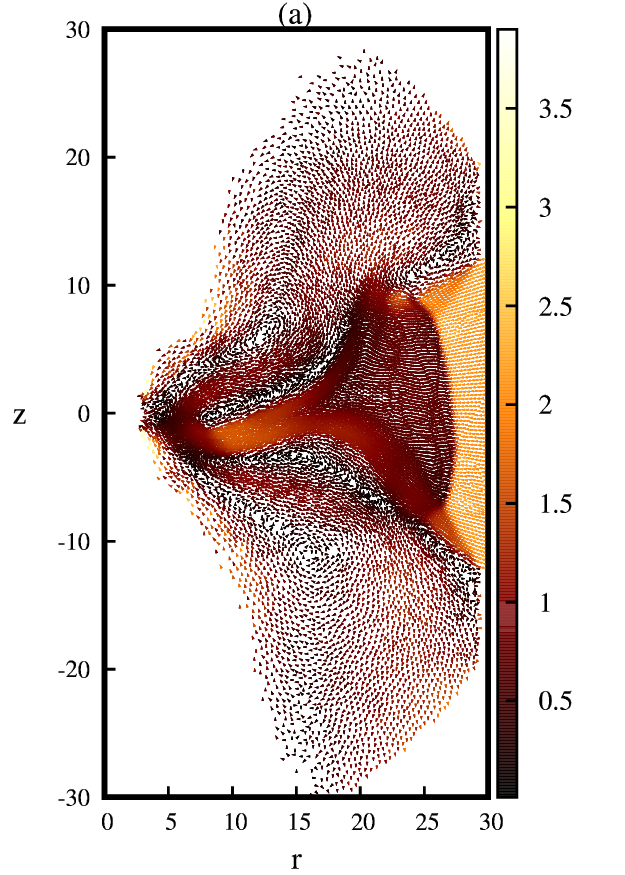}
\includegraphics[height=5.0cm,width=3.0cm]{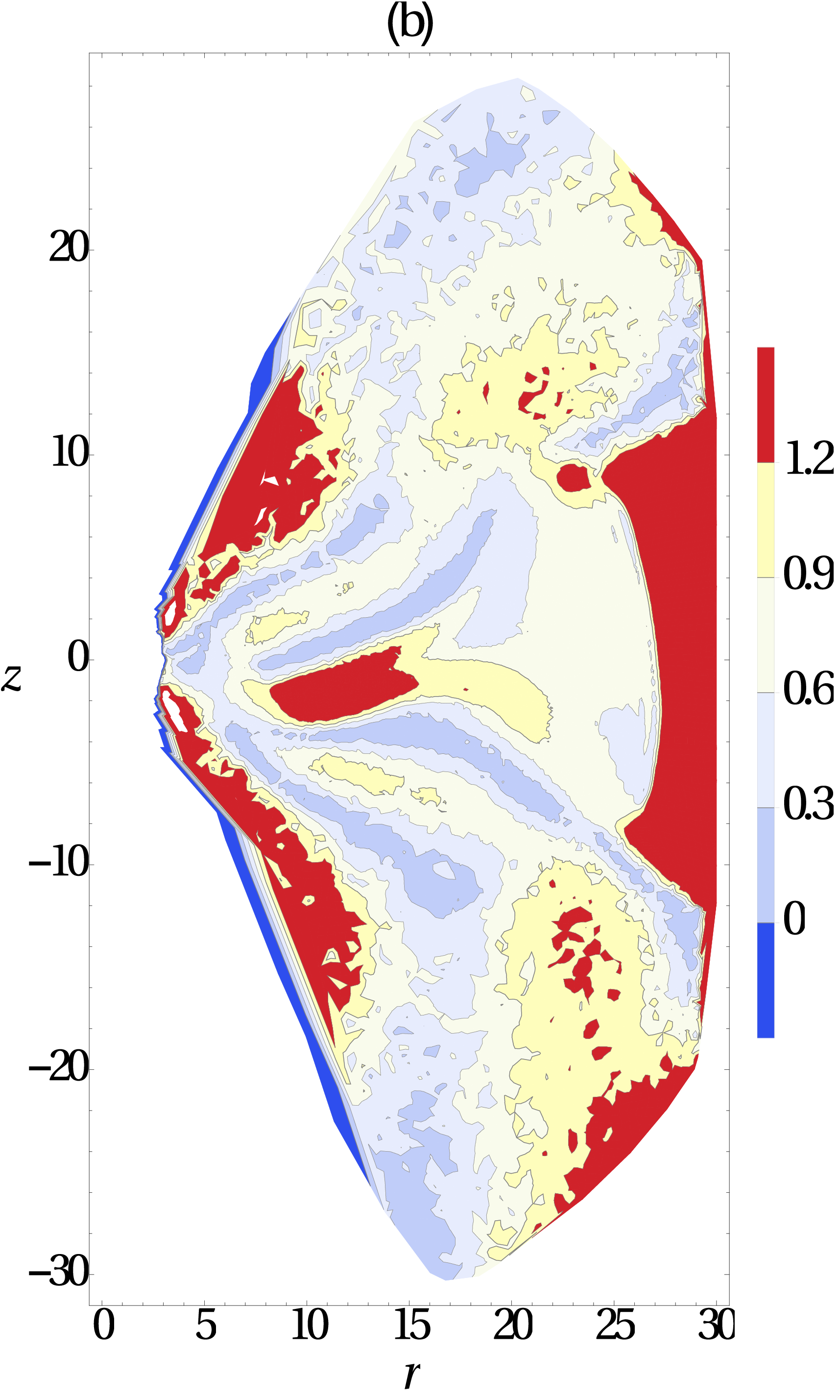}
	\caption{\footnotesize{(a) Velocity vector $v_r \hat{r} + v_z \hat{z}$ (arrow heads) with Mach number in the colour bar, for the flow configuration $C1$ at time $t=0.2800 s$ and (b) corresponding contours of constant Mach number range showing the outer and inner shocks.}}
\end{figure}

\textbf{Case C2:} When the viscosity parameter is increased from $0.075$ to $0.15$, the size of the NBOL is increased from $\sim 2 r_S$ to $\sim 5.5 r_S$. The size of RAKED is also increased from $\sim 2.5 r_S$ in C1 to $\sim 3.5 r_S$ here, though it has relatively weaker oscillations. The outer part of the flow remained sub-Keplerian throughout the simulation. This case is of particular interest as all the layers are present simultaneously. When the velocity vectors and Mach numbers are compared with those in C1, we notice that the flow has become more stable due to introduction of viscosity and consequent decrease of specific angular momentum in the inner regions. The outer shock is moved at a larger distance from the star boundary. The NBOL did not eject any matter due to the lack of centrifugal drive by low angular momentum and only the CENBOL ejected matter. A part of the outflowing matter is seen to move towards the NS surface transonically. The flow interacts with a larger area on the NS surface due to the further reduction of angular momentum near the star. The inner shock location moves closer to the NS and is narrower.
\begin{figure}[h]
\centering
\includegraphics[height=5.0cm,width=3.0cm]{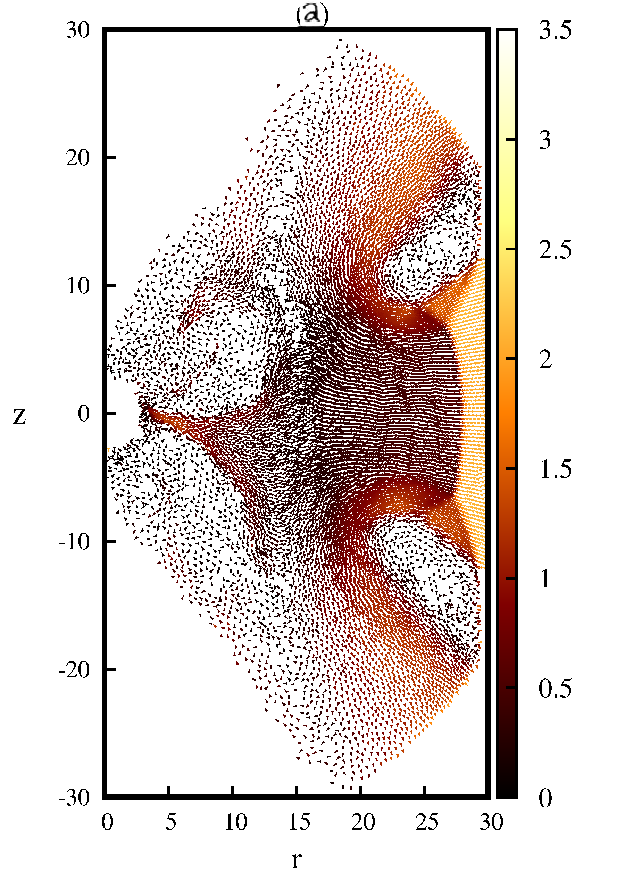}
\includegraphics[height=5.0cm,width=3.0cm]{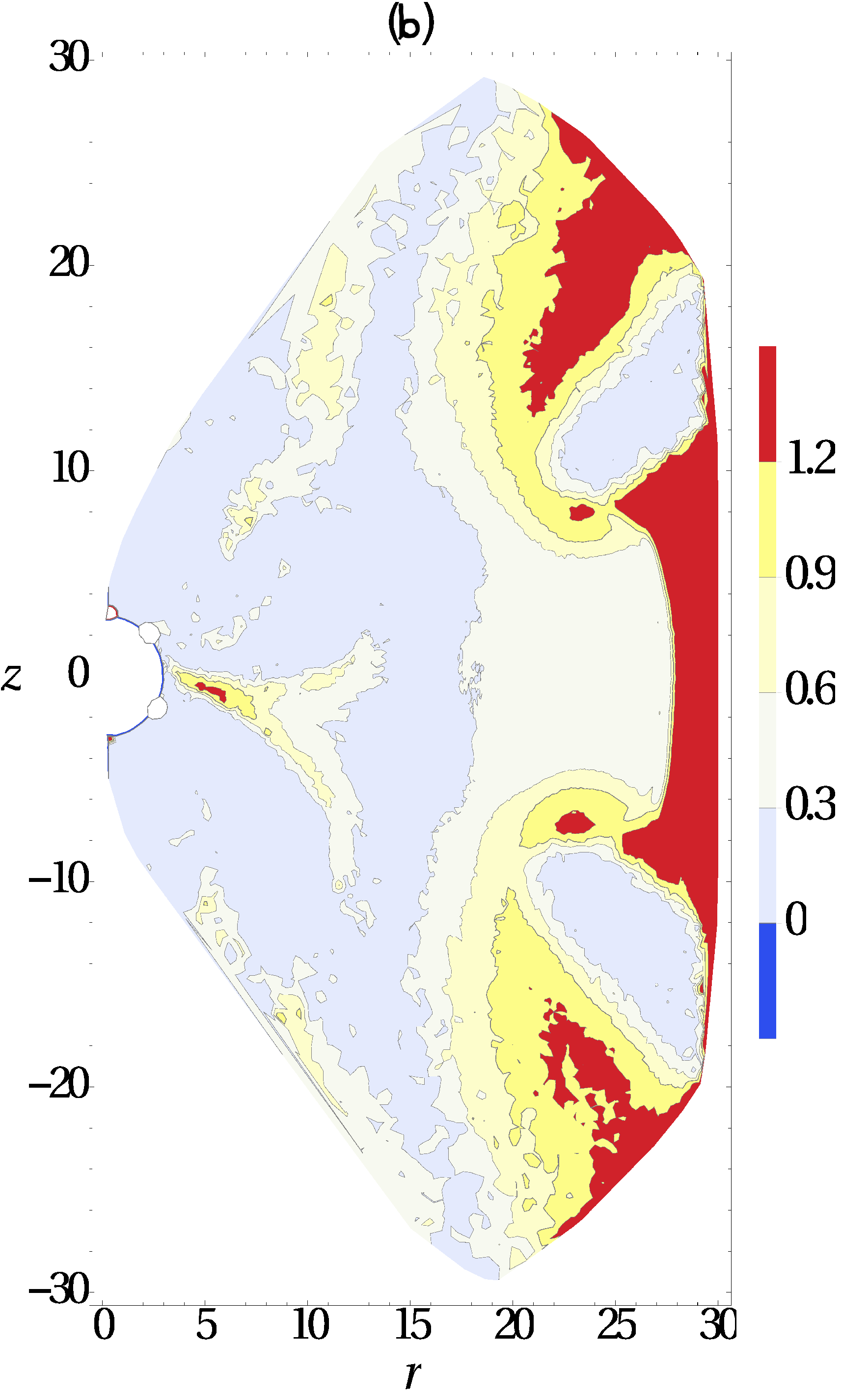}
	\caption{\footnotesize{(a) Velocity vector $v_r \hat{r} + v_z \hat{z}$ (arrow heads) with  Mach number in the colour bar, for the flow configuration $C2$ at time $t=0.1400 s$ and (b) corresponding contours of constant Mach number range showing the outer and inner shocks.}}
\end{figure}

\textbf{Case C3:} Further increase of viscosity parameter from 0.15 to 0.3 leads to increase of NBOL size to $\sim 11.5 r_S$. Furthermore, the RAKED kept growing in size as its outer boundary started to increase with increasing time. The inner edge of RAKED (outer edge of NBOL) remained fairly constant. We also show the density, temperature and Mach number contours of this case at time $t=0.04 s$, in Figs. 3(a), 3(b) and 3(c), respectively. Both the density jumps near CENBOL and NBOL are distinctly seen in 3(a) and 3(c). We also notice that matter is ejected from near the the outer shock. The temperature distribution of the flow also captures the shock transition. In addition to those, further hotter and clumpy regions are seen near the NBOL where the inflow and outflows mix. In this case, the shock is pushed to the outer boundary very quickly and matter is ejected in both quadrants from CENBOL. NBOL is almost symmetric about $z=0$ and did not produce any outflow. Fallback of some matter onto the star is also seen.
\begin{figure}[h]
\centering
\includegraphics[height=3.5cm,width=2.0cm]{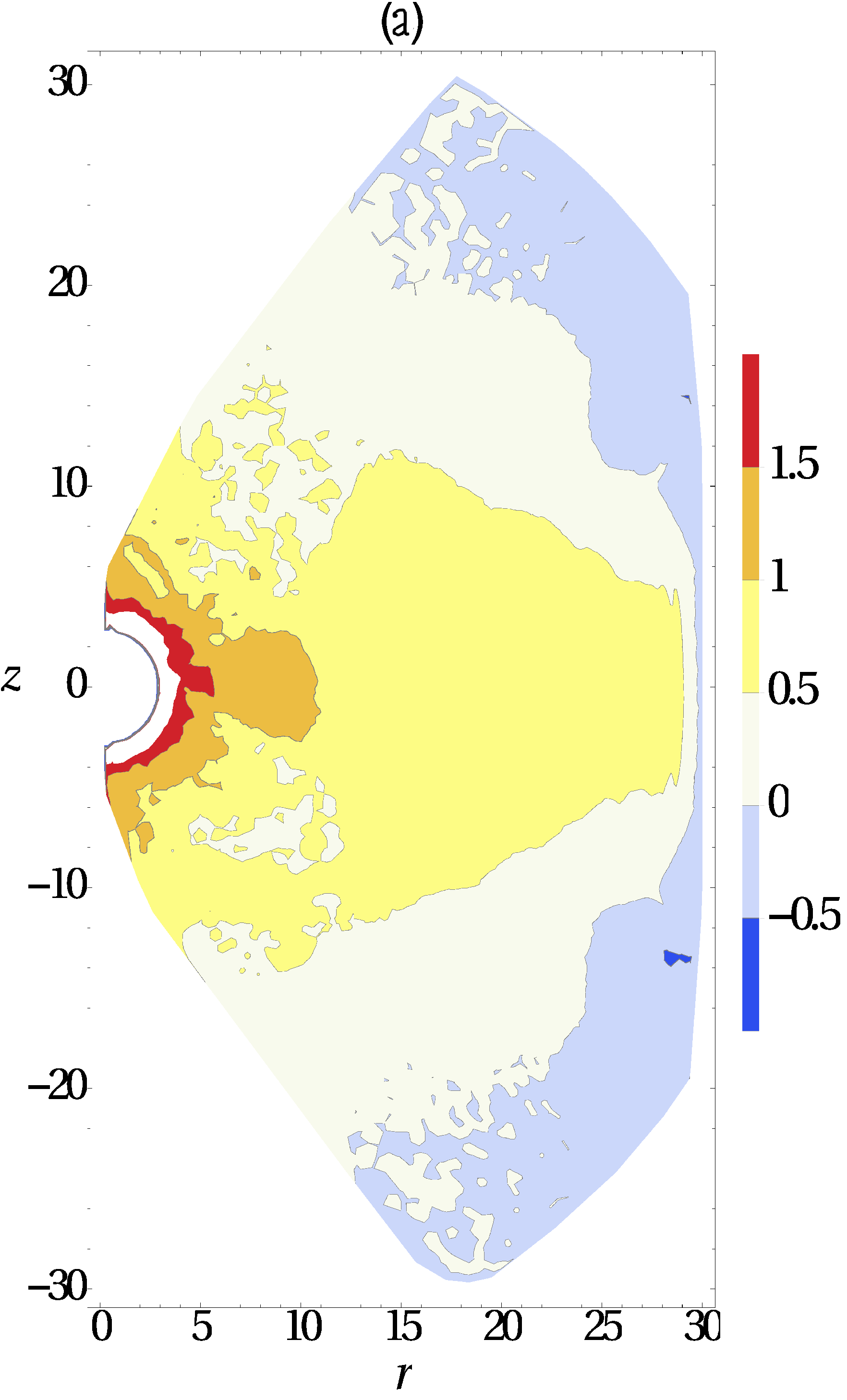}
\includegraphics[height=3.5cm,width=2.0cm]{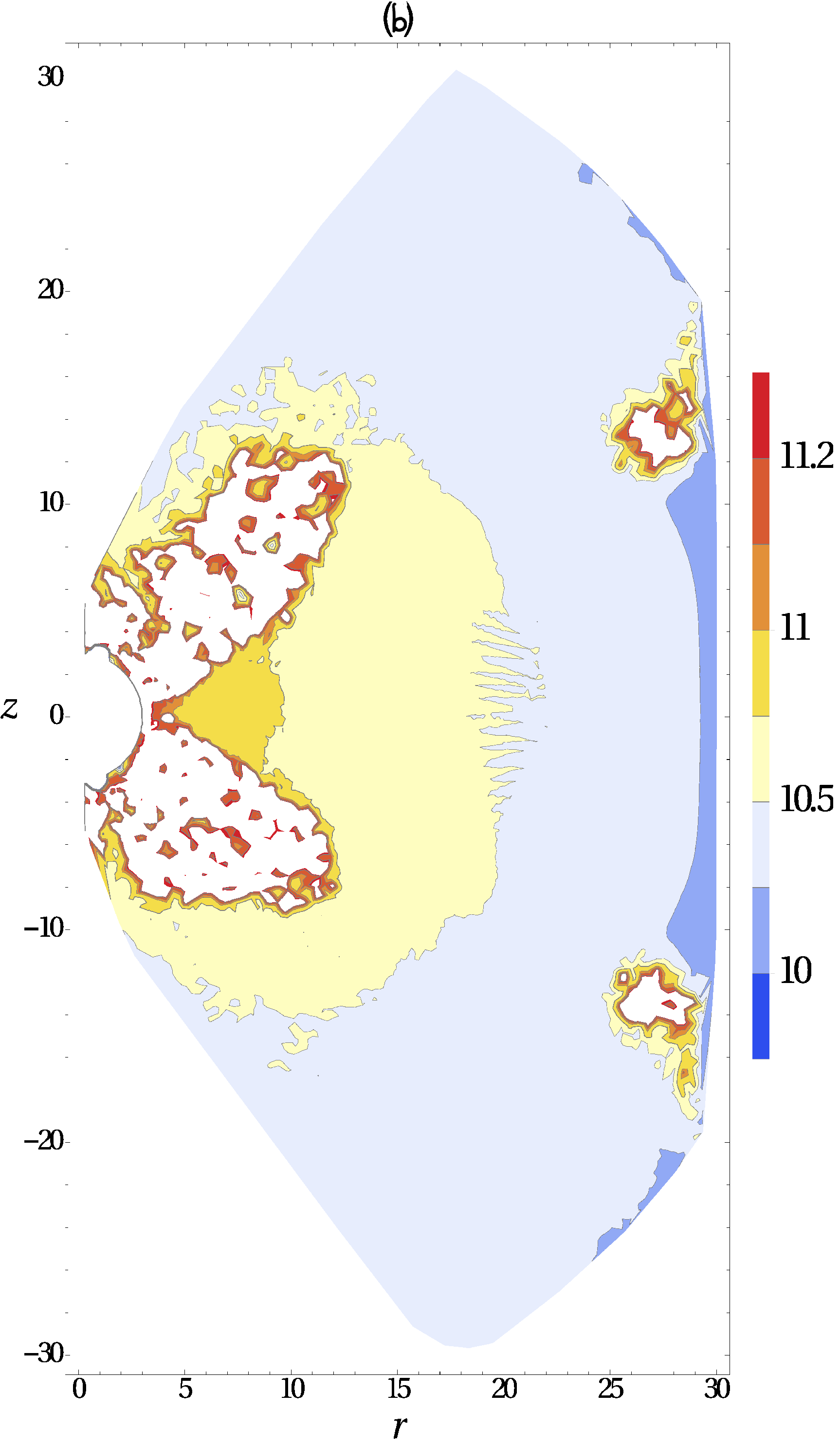}
\includegraphics[height=3.5cm,width=2.0cm]{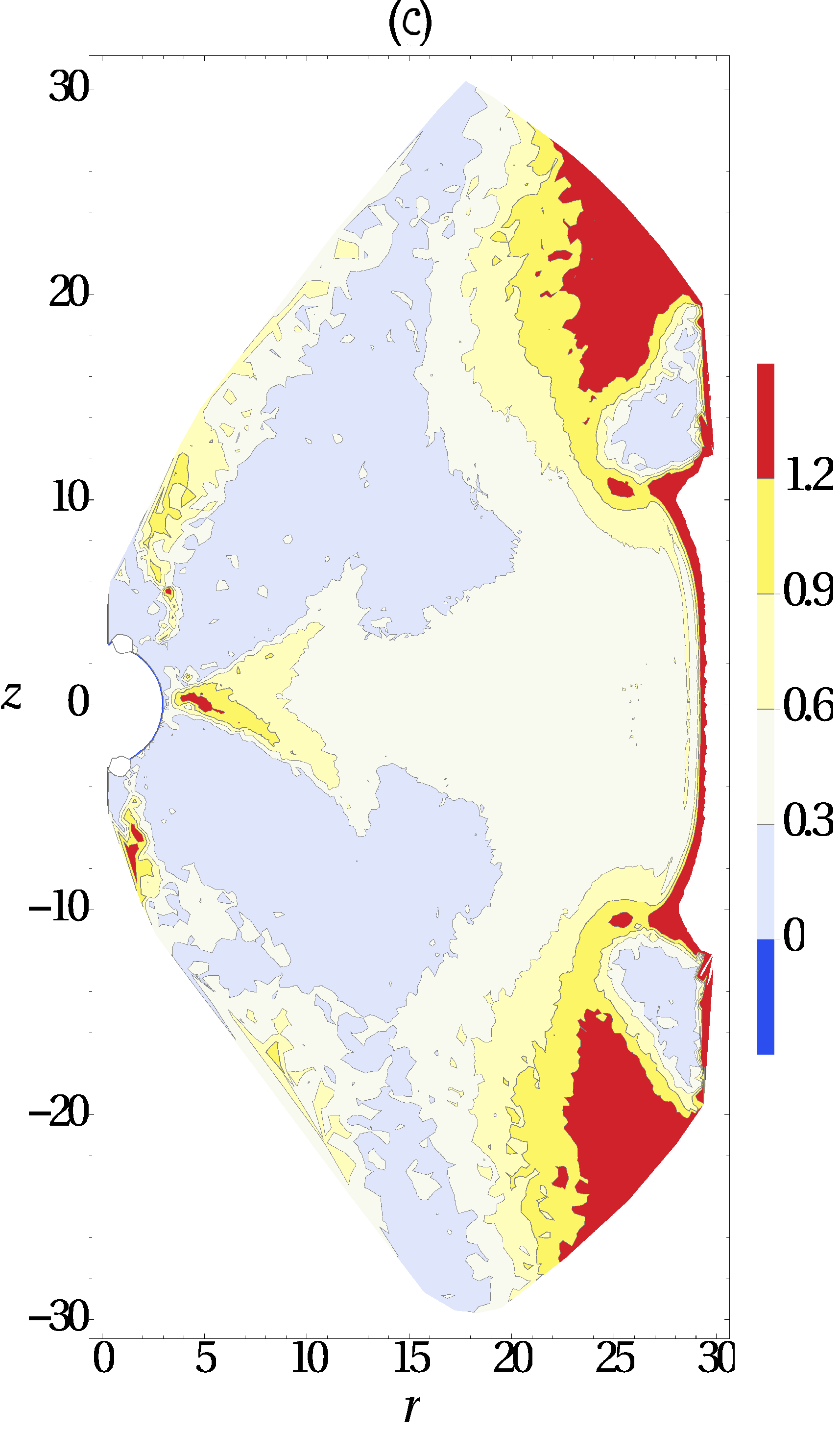}
\caption{\footnotesize{(a) Contours of constant $log(\rho/\rho_0)$ range for case $C3$ at time $t=0.0400 s$ and (b) Corresponding contours of constant temperature range in $K$ (log scale) and (c) Corresponding Mach number range contours.}}
\end{figure}
\begin{figure}[h]
\centering
\includegraphics[height=5.0cm,width=3.0cm]{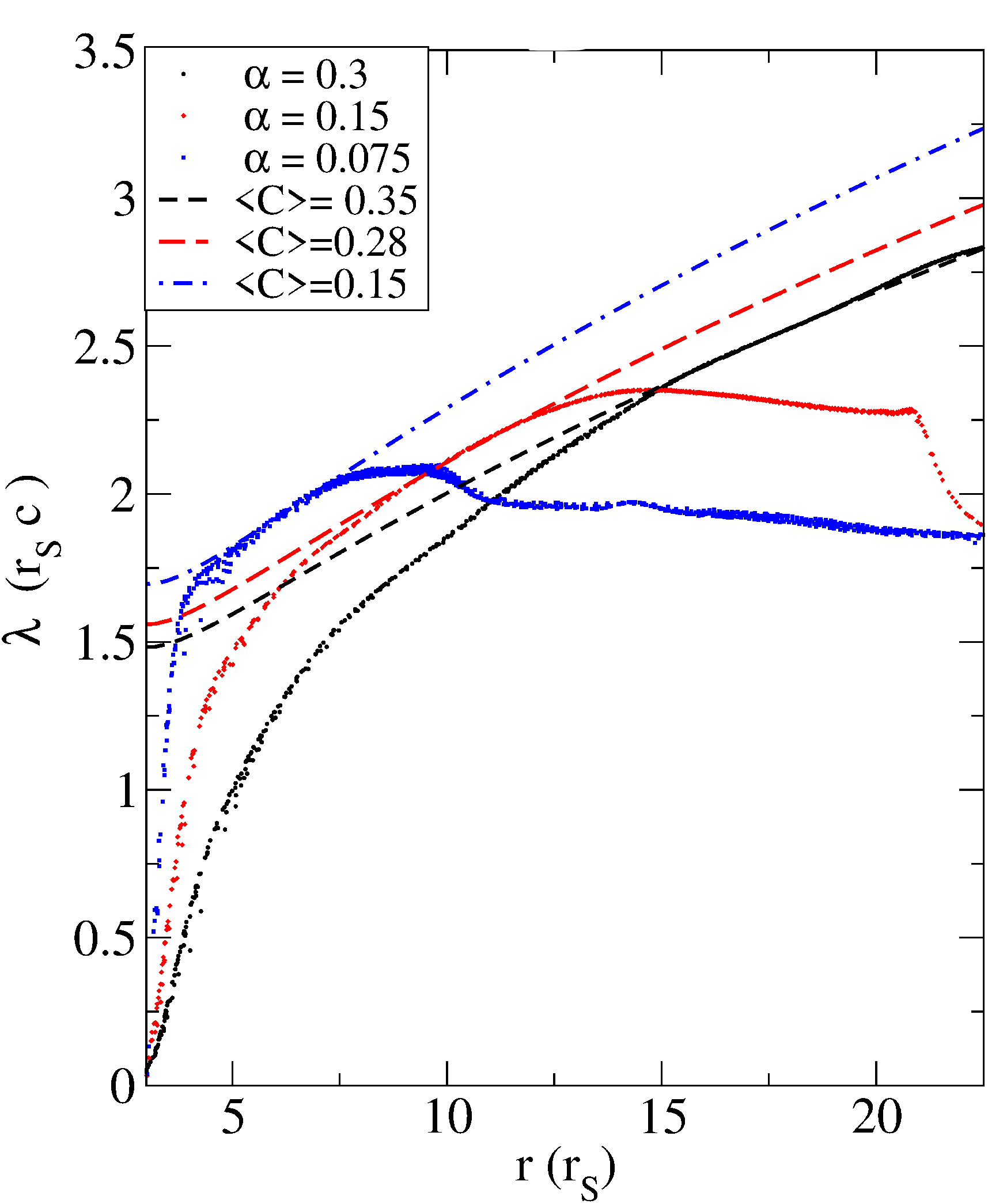}
\caption{\footnotesize{Formation of different regions in inflow for various $\alpha$, at a time $t=0.016 s$. Dashed curves are the theoretical distribution with $\frac{\sqrt{1-<C>}r^{1/2}}{\sqrt{2}(1-1/r)}$ shape, while the dotted curves are as obtained from the simulations.}}
\end{figure}
\begin{table}
\caption{\footnotesize{Outer radius of different regions of the inflow for different values of $\alpha$, at $t=0.016~s$, for $-0.5<z<0.5$.}} 
\centering
\tiny{
\begin{tabular}{c c c c c c c}
\hline
\hline
No. & $\alpha$ & $<C>$ & NBOL & RAKED & CENBOL \\
\hline
\hline
C1 & 0.075 & 0.15 & 5.0 & 7.5 & 17.5\\
\hline
C2 & 0.15 & 0.28 & 8.5 & 12.0 & 21.0\\
\hline
C3 & 0.3 & 0.35 & 14.5 & 23.5 & 27.5\\
\hline
\hline
\end{tabular}
}
\end{table}
The $\lambda$ vs $r$ curve is drawn in Fig. 4 to show the region where Raked is formed. The dashed curves are theoretical shapes $\lambda = \frac{\sqrt{1-<C>}r^{1/2}}{\sqrt{2}(1-1/r)}$ while the dotted curves are as obtained from simulations. The values of $<C>$ increased with the increase of $\alpha$, suggesting that more energetic matter (due to viscous heating) makes its way to the NS surface and raised the radiative pressure exerted on the inflowing matter.
\subsection{Formation of Two-Component Advective Flows}
We introduce a vertical variation in the viscosity parameter $\alpha\sim\alpha_0 exp(-|z|)$, where $\alpha_0=0.3$. This ensures that the viscosity is highest at the equatorial plane and sharply falls off with distance from the equatorial plane. When such a differential viscosity is introduced, a viscous disk appears to form on the equatorial plane and the matter away from it continues to advect with sub-Keplerian angular momentum. In Fig. 5(b), the contours show the formation of outer shock near $20~r_S$ on both quadrants. The inflow undergoes another set of shocks near the surface of the star between $3-5~r_S$. The region along the equatorial plane has low mach number $(0.3-0.6)$ and interacts with the outer shock. Near the inner region, both flows undergo some mixing, which leads to a fluctuation in the angular momentum distribution as seen in Fig. 5(a).
\begin{figure}[h]
\centering
\includegraphics[height=5.0cm,width=3.0cm]{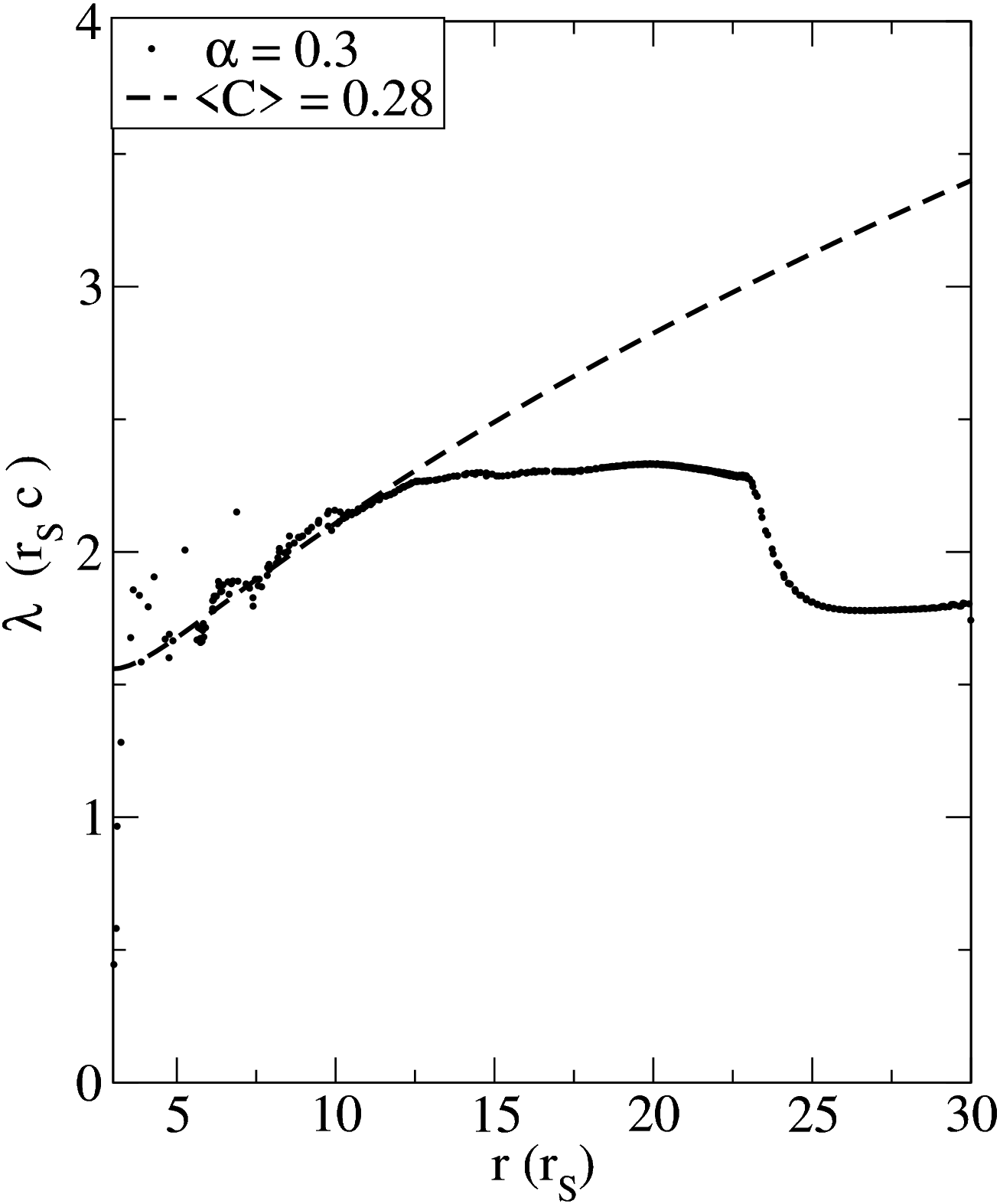}
\includegraphics[height=5.0cm,width=3.0cm]{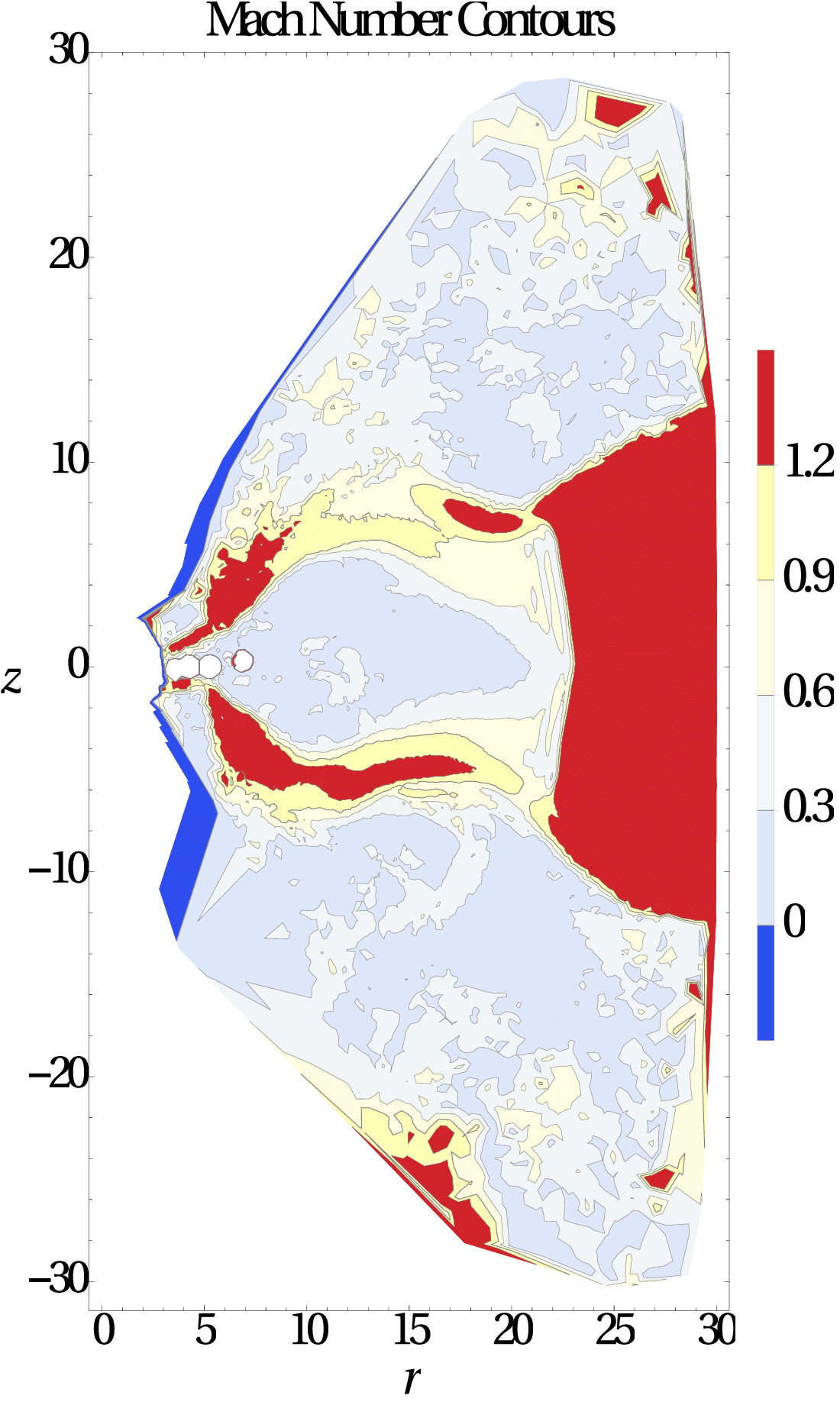}
\caption{\footnotesize{Panel 1:Formation of different regions in inflow for a differential $\alpha$, at a time $t=0.12 s$. Dashed curves are the theoretical distribution with $\frac{\sqrt{1-<C>}r^{1/2}}{\sqrt{2}(1-1/r)}$ shape, while the dotted curves are as obtained from the simulations. Panel 2: Mach Number contours for the corresponding case, showing the shocked flows (sub-Keplerian) above and below the viscous region where Mach Number is low, as the flow is Keplerian like, denoting the formation of TCAF around neutron stars.}}
\end{figure}

\section{Conclusions}
In this paper, we report the formation of a normal boundary layer and a radiative Keplerian disk out of a sub-Keplerian injected flow around a weakly magnetic neutron star. To our knowledge, this is the first detailed study of viscous flows around a weakly magnetized neutron star. We carried out a qualitative study of the behaviour of three regions formed in the inflowing matter as the flow viscosity is gradually increased. When viscosity is low, resembling an inviscid flow, matter adjusts its angular momenta only very close to the NS surface and forms the NBOL out of the sub-Keplerian flow. RAKED is transient in nature for these cases. We believe this happens in HMXBs, such as, Cir X-1. We notice that two regions, one on the inner edge of CENBOL and the other near NBOL, eject matter along the z-direction. A part of these two matter merge and a part actually falls back to these regions. For higher viscosity, the RAKED is formed in between NBOL and the sub-Keplerian flow. Here, the RAKED is small and steadily oscillate. These would be the most general type of inflow in presence of viscosity. We also notice that the redistribution of angular momentum helps in ejecting more matter out of the equatorial plane, even more compared to inviscid cases. For even higher viscosity, a clear RAKED is formed and increased in size towards larger radial distance. This suggests that these cases are possible in the super-critical range of viscosity. Within 25 dynamical timescales at $30 r_S$, the RAKED reached the outer edge. The redistributed angular momentum also leads to an even higher ejection of matter from the disk. The added viscosity appears to make the flow more stable and the vertical oscillations become negligible. The NBOL in Case C3 is equivalent of the transition layer of Titarchuk, Lapidus and Muslimov (1998). However, our simulations generate the Keplerian disk self-consistently, compared to their cases where it was assumed to be Keplerian. In addition to that, we find sub-Keplerian to sub-Keplerian transition layers in Case C2, which was not reported before. Case C1 is a equivalent to the spreading boundary layer as the layer is very thin and close to the surface. Here, too, the assumption of Keplerian disk was not used by us and thus our simulations present a more generalized solution. We also find that due to the high density of matter near the surface, the inflow is more likely to form the NBOL type layer first, before a disk can be formed near the surface. This also connects all the three approaches by the variation of viscosity in the flow. When a differential viscosity is introduced through a Z-gradient, the equatorial region with high viscosity forms a disk and the region away from it, maintains sub-Keplerian angular momentum in the inflow. This leads to the formation of a TCAF geometry.  
\acknowledgments
\footnotesize{
AB acknowledges the computational support provided by SNBNCBS. SKC acknowledges support from the DST/SERB sponsored Extra Mural Research project (EMR/2016/003918) fund.
}

\end{document}